\documentstyle[amssymb,multicol,prb,aps]{revtex}

\begin{document}
\draft
\title{NMR relaxation rates and Knight shifts in MgB$_2$}
\author{Eva Pavarini }
\address{INFM-Dipartimento di Fisica ``A.Volta'', 
Universit\`a di Pavia, I-27100 Pavia, Italy}
\author{I.I. Mazin}
\address{
Code 6390, Naval Research Laboratory, Washington, DC 20375}
\date{\today}
\maketitle

\begin{abstract}
We calculate {\em ab initio} the NMR relaxation rates and the Knight shifts
in MgB$_{2}$. We show that the dominant relaxation mechanism at the $^{11}$B
nucleus is the interaction with the electronic orbital moment, and we give a
simple explanation of that using a simple $sp$ tight binding model. When
Stoner enhancement (also calculated {\em ab-initio}) is accounted for, we
obtain good agreement with reported experimental values. For the $^{25}$Mg
nucleus, we predict that the dominant relaxation mechanism is the
Fermi-contact interaction, which also dominates the Mg Knight shift.
\end{abstract}


 \begin{multicols}{2}

Recent discovery\cite{Nagamatsu} of superconductivity in MgB$_{2}$ created
substantial interest. It was suggested that the underlying mechanism is
electron-phonon interaction in the boron sublattice\cite{Igor}, which was
subsequently confirmed by observation of a sizeable boron\cite{Budko,Mg},
but not magnesium\cite{Mg} isotope effect. State of the art local density
approximation (LDA) calculations\cite{Kong,heid,multigap} produced
electron-phonon coupling constants $\lambda $ ranging from 0.75 to 0.87.
Lacking single crystals, experimental determination of $\lambda $ relies on
the specific heat renormalization measurements\cite{sheat}. Using the LDA
density of states (DOS), these experiments give $\lambda \sim 0.6-0.8$; however,
if there is any many-body renormalization of the LDA DOS,
these experiments should be reanalyzed.

Nuclear magnetic resonance (NMR) is a common probe of the DOS.
The measured quantities, the spin-lattice relaxation rate, $1/T_{1},$ and
the Knight shift, $K,$ are related to the spin susceptibility, and thus are
not subject to a phonon renormalization. Measurements of the relaxation
rates and the Knight shift of $^{11}$B already exist\cite%
{nmrames,Tou,nmrrussia,nmrosaka}. From the electronic structure of MgB$_{2}$
one can conjecture\cite{nmrames,Tou} that the main source of relaxation
should be the hyperfine coupling between the nuclear spin and conduction B p
electrons. However, a full microscopic understanding of the NMR data
(relaxation rates and Knight shifts) is still missing. While different
sources\cite{nmrames,nmrrussia,nmrosaka} reasonably agree among themselves
about the relaxation rates, reporting $1/T_{1}T$ between 5.6$\times 10^{-3}$
and 6.5$\times 10^{-3}$ 1/(K sec), there is considerable controversy about
the Knight shifts. Some authors\cite{nmrrussia} report a small average shift 
$K=(K_{z}+2K_{xy})/3=0.0175\%$, and give an upper bound on its anisotropy, $%
K_{ax}=(K_{z}-K_{xy})/3<0.0030\%$. Other authors\cite{nmrames} report even
smaller ($K=0.006\%$) shift and they attribute the shift to the
Fermi-contact interaction. Note that the Korringa relation,  $%
r=K^{2}(T_{1}T)(\gamma _{n}/2\mu_B)^{2}(4\pi k_{B}\hbar )\approx 1$,
where $\gamma _{n}$ is the nuclear gyromagnetic
ratio, is not satisfied here, as the measurements give $
r\sim 0.2$. Finally, a tiny negative
shift ($K=-0.0005\%$) was measured by Tou et al.,\cite{Tou} and attributed
to core polarization. These discrepancies might arise from the difficulties
in measuring the $^{11}$B shift, due to its smallness, and, possibly, from
the selection of the reference material.\cite{nmrames,Tou} Therefore, in
order to clarify the microscopic origin of the NMR relaxation process and of
the Knight shift, {\em ab-initio} calculations are highly desirable.

In the present work we report LDA calculation of the relaxation rates and of
the Knight shifts. We will show that for $^{11}$B the relaxation is due to
the $p$ states, and the {\em orbital} relaxation rate is about 3 times
larger than the {\em dipole} rate and 10 time larger than the {\em %
Fermi-contact} rate. After an appropriate Stoner renormalization is
included, the agreement with the experiment is very good. 
On the other hand, the main source of Knight shift is the hyperfine coupling
with $s$ electrons. Also, the (yet unmeasured) relaxation on Mg is mainly
due to the Fermi-contact interaction with the $s$ states.

The hyperfine interaction $-\hbar\gamma_n{\bf I} \cdot {\bf H}$ is the
coupling between the nuclear magnetic moment $\hbar\gamma_n{\bf I}$ and the
hyperfine field {\bf H} produced at the site of the nucleus by the
conduction electrons. In order to discuss separately the different
relaxation mechanisms, we neglect the small spin orbit coupling and split
the hyperfine interaction into three terms, $-\hbar\gamma_n {\bf I}\cdot[ 
{\bf H}^{o}+{\bf H}^{d}+{\bf H}^{F}]$. The first term is the coupling with
the electronic orbital moment; the second and the third terms are,
respectively, the dipole and the Fermi-contact interaction with the
electronic spin. Thus the total hyperfine field is given by 
\[
{\bf H}=2\mu_B \left\{ -\frac{{\bf l}}{r}+\left[ \frac{{\bf s}}{%
r^{3}}-3\frac{{\bf r}({\bf r}\cdot {\bf s})}{r^{5}}\right] -\frac{8\pi {\bf s%
}}{3}\delta ({\bf r})\right\} , 
\]%
where {\bf r}, {\bf s} and {\bf l} are the electronic position, spin, and
angular momentum operator. In the case of $^{11}$B, $I=3/2$ and $\gamma
_{n}=0.89\;\gamma _{N}$, while in the case of $^{25}$Mg $I=5/2$ and $\gamma
_{n}=-0.17\gamma _{N}$, with $\gamma _{N}=e/m_{p}c$.


According to Fermi's golden rule, the relaxation rate, $1/T_{1}$, may be
written as\cite{nmr}
\begin{eqnarray}
&&\frac{1}{T_{1}}=\frac{2\pi }{\hbar }\sum_{{\bf kk}^{\prime }{ss}^{\prime }{%
mm}^{\prime }}f(\epsilon _{{\bf k}s})[1-f(\epsilon _{{\bf k}s^{\prime
}})]\delta (\epsilon _{{\bf k}s}-\epsilon _{{\bf k}^{\prime }s^{\prime }}) 
\nonumber \\
&\times &|\langle {\bf k}sm|\!-\!\hbar \gamma _{n}{\bf I}\cdot {\bf H}|{\bf k%
}^{\prime }s^{\prime }m^{\prime }\rangle |^{2}\frac{\left( {\langle }%
m|I_{z}|m{\rangle \!}-\!\langle m^{\prime }|I_{z}|m^{\prime }{\rangle }%
\right) ^{2}}{{\rm Tr}I_{z}^{2}},  \label{eq3}
\end{eqnarray}%
where $f(\epsilon )$ is the Fermi-Dirac distribution, {\it s} is the spin
index, and $|m\rangle $ are the eigenstates of $I_{z}$. Here ${\bf k}$
stands for both the wave vector and band index. Expansion of the Fermi
function and integration over the nuclear spin yields, for a polycrystalline
sample, the following expression\cite{Antropov} 
\[
\frac{1}{T_{1}T}={2\pi k_{B}\hbar}\gamma _{n}^2
\left[ {\rm Tr}\frac{1}{3}|{\bf H}N|^{2}\right] ,
\]%
where $N$ is understood as a diagonal matrix in the spin and {\bf k}-space, $%
\delta _{ss^{\prime }}\delta _{{\bf kk}^{\prime }}\delta (\epsilon _{{\bf k}%
})$. 
The relevant  prefactor is 
$C=({4\pi k_{B}/\hbar })\left( {\gamma }_{n}/\gamma
_{e}\right) ^{2}$, the same that appears in the Korringa relation. In the
present case $C\sim 1.4\times 10^{4}$/(K sec) for $^{25}$Mg and $C\sim
3.9\times 10^{5}$ /(K sec) for $^{11}$B.~The interaction cross terms in Eq.\
(\ref{eq3}), i.e. the terms proportional to Tr[{\bf H}$^{o}N${\bf H}$^{d}N$%
], Tr[{\bf H}$^{o}N${\bf H}$^{F}N$] and Tr[{\bf H}$^{F}N${\bf H}$^{d}N$] all
vanish, the first two exactly, because Tr[{\bf s}]=0 and the third vanishes
exactly for polycrystals because Tr[$s^2-3({\bf s\cdot r})^2]=0$, and
 approximately for single crystals, when the $d$-electron  DOS
is small({\it cf.} Ref. 
\onlinecite{asada-hcp}, Eq. 24). Thus, without the core
polarization, which will be discussed later,  the relaxation rate has
three contributions: the orbital, the dipole and the
contact-field term. Note that in the terminology of Ref.\cite{asada-hcp},
all cross terms, diagonal in interaction but off-diagonal in angular
momentum, are included in the calculation. More details on this derivation
can be found in Ref.\onlinecite{Antropov}.

In order to evaluate the relaxation rate, we adopt the tight binding
LMTO-ASA method (LMTO47 Stuttgart code).\cite{lmto} This method has been
already used with success to calculate $1/T_{1}$, e.g.~in~A$_{3}$C$_{60}$.%
\cite{Antropov} Thus we express the Bloch function as $|i{\bf k}s\rangle
=\sum_{RL}\langle {\bf r}|{\chi }_{RL}^{{\bf k}}\rangle c_{RLi,{\bf k}%
}|s\rangle $, with $|{\chi }_{RL}^{{\bf k}}\rangle =|\Phi _{RL}\rangle
+\sum_{R^{\prime }L^{\prime }}|\dot{\Phi}_{R^{\prime }L^{\prime }}\rangle
h_{R^{\prime }L^{\prime },RL}^{{\bf k}}$.~Here~$\langle r|\Phi _{RL}\rangle
=\phi _{Rl}(\epsilon _{\nu Rl},r)Y_{L}(\hat{{\bf r}}_{R})$, where $\phi _{Rl}
$ is the radial solution of the Schr\"{o}dinger equation at the energy $%
\epsilon _{\nu RL}$, $\dot{\phi}_{Rl^{\prime }}$ is its energy derivative, $%
Y_{L}$ is a spherical harmonic with $L=lm$. For simplicity, in the following
we will write only the contributions from $\phi _{Rl}$, although in the
calculation we have, of course, included all terms. Thus the three
contributions to $1/T_{1}$ can be expressed as 
\begin{equation}
N_{LL^{\prime }}=\frac{V}{8\pi ^{3}}\sum_{i}\int dk^{3}c_{L,i{\bf k}}\delta
(\epsilon _{i{\bf k}})c_{L^{\prime },i{\bf k}}^{\ast }\;,
\end{equation}%
and of the radial integrals involving $\phi _{Rl}(\epsilon _{\nu Rl},r)$ 
\begin{equation}
\langle r^{-3}\rangle _{l^{\prime }l}=\int \phi _{Rl}(\epsilon _{\nu
Rl},r)r^{-3}\phi _{Rl^{\prime }}(\epsilon _{\nu Rl^{\prime }},r)r^{2}dr.
\end{equation}%
The Fermi-contact, the orbital, and the dipole contributions may then be
written respectively as 
\begin{eqnarray}
Tr{\frac{1}{3}}|{\bf H}^{F}N|^{2} &=&{\frac{1}{2}}{\mu _{B}^{2}}\left( {%
\frac{4}{3}}\phi _{s}^{2}(\epsilon _{\nu Rl},0)N_{ss}\right) ^{2}, \\
\ Tr{\frac{1}{3}}|{\bf H}^{0}N|^{2} &=&{\frac{8}{3}}\mu _{B}^{2}\sum_{\mu
=-1}^{1}\sum_{\Lambda \Lambda ^{\prime }LL^{\prime }}  \nonumber \\
&\times &\langle r^{-3}\rangle _{\lambda \lambda }D_{LL^{\prime }}^{-\mu
}N_{L^{\prime }\Lambda ^{\prime }}\langle r^{-3}\rangle _{ll}D_{\Lambda
^{\prime }\Lambda }^{\mu }N_{\Lambda L}, \\
Tr{\frac{1}{3}}|{\bf H}^{d}N|^{2} &=&{4}\mu _{B}^{2}\sum_{\mu
=-2}^{2}\sum_{\Lambda \Lambda ^{\prime }LL^{\prime }}  \nonumber \\
&\times &\langle r^{-3}\rangle _{\lambda ^{\prime }\lambda }C_{LL^{\prime
}}^{2\mu }N_{L^{\prime }\Lambda ^{\prime }}\langle r^{-3}\rangle
_{ll^{\prime }}C_{\Lambda \Lambda ^{\prime }}^{2\mu }N_{\Lambda L}.
\end{eqnarray}%
Here $D_{LL^{\prime }}^{\mu }=\langle L^{\prime }|l_{\mu }|L\rangle $, $%
l_{0}=l_{z},$ $l_{\pm 1}=l_{\pm }/\sqrt{2}$, and $C_{LL^{\prime }}^{2\mu }=%
\sqrt{\frac{4\pi }{5}}\int Y_{2\mu }(\hat{r})Y_{L}(\hat{r})^{\ast
}Y_{L^{\prime }}(\hat{r})d^{2}\hat{r}$


In the same way, the Knight shift can be written as $K_{\alpha
}=2\mu_B\;Tr\langle \uparrow |{\bf H}{_{\alpha }}N|\uparrow \rangle $ where $%
\alpha $ is the direction along which the external magnetic field is
applied. As the relaxation rate, the relative shift may also be expressed as
a function of the DOS matrix and the radial integrals,
expanding the Bloch function in the LMTO basis set.

The DOS matrix was calculated by the linear tetrahedron
method. We found that the results were already very well converged with a
mesh of 370 irreducible {\bf k} points. In order to minimize the
linearization error accurate wavefunctions at the Fermi level, the final
runs were performed with $\epsilon _{\nu Rl}=\epsilon _{F}$. The convergence
of the sums over the angular momentum was also very good. We find that we
can truncate after $l=2$. The reason is that the radial integrals $\langle
(a_{0}/r)^{3}\rangle _{ll^{\prime }}$ ($a_{0}$ is the Bohr radius) decrease
quickly when $l$ and $l^{\prime }$ increases. For Mg we find, e.g. $\langle
(a_{0}/r)^{3}\rangle _{11}=4.8$, $\langle (a_{0}/r)^{3}\rangle _{22}=0.16$,
and $\langle (a_{0}/r)^{3}\rangle _{33}=0.09$, and in for B, $\langle
(a_{0}/r)^{3}\rangle _{11}=1.1$ and $\langle (a_{0}/r)^{3}\rangle _{22}=0.2$.

What is the dominant mechanism that gives rise to the magnetic relaxation at
B and Mg nuclei? In most metals it is the Fermi contact
one, defined by the DOS of the $s$ electrons at the
Fermi level. 
However, in the case of MgB$_{2}$ the states near the Fermi level are mainly
B p. We find that the ratio $N_{ss}({\rm Mg})/N_{tot}({\rm Mg)}\sim 1/4,$
and $N_{ss}({\rm B})/N_{tot}({\rm B)}\sim 1/50$. Therefore, at least in the
case of B, the ratio is very small, and the Fermi contact term could become
comparable or even smaller than the dipole or the orbital term. We have
calculated all three contributions for both elements and show the results in
Tables 1 and 2.

We also calculate {\em ab initio} the core polarization. For this purpose we
applied in the calculations an external magnetic field $B$, and then
calculated $m_{n}(0)$, the spin density of the $n$th core shell at the
nucleus. Then the core polarization Knight shift can be obtained as $%
K_{cp}=\mu _{B}(8\pi /3)\sum_{n}(m_{n}(0)/B)$, and the corresponding
contributions to the relaxations rate can be computed from the Korringa
relations for the core states\cite{cp}. For $^{25}$Mg, the contribution of
the core polarization is negligible and the Fermi-contact interaction
dominates. For $^{11}B,$ we find that the contribution of the $1s$ shell is
of the same order, but of opposite sign, as the the $2s$ shell contribution.
Their total effect is thus small. Also, since the Fermi-contact contribution
for B is much smaller than in Mg, the relative effect of the dipole term is
larger, leading to a noticeable anisotropy of the Knight shift (about 30\%),
while the Mg Knight shift is essentially isotropic.


\end{multicols}
\begin{table}[tbp]
\caption[Tab. 1]{Knight shift, $K_{\protect}\protect\alpha $ in \%. Both
unrenormalized and Stoner-enhanced values are included, as discussed in the
text. The label $\protect\alpha =$ab,c indicates the direction of the
external magnetic field. }
\begin{tabular}{c|cccccccccc}
& dipole (ab) & dipole (c) & orbital & {\ Fermi-contact} & {core} & Total
(ab/c) & Total (renormalized) & Expt.\tablenote{Ref.\onlinecite{nmrrussia};
$^{\rm b}$ Ref.\onlinecite{nmrames}; ${\rm ^c}$ Ref.\onlinecite{Tou}}
& Expt. $^{\rm b}$ & Expt. ${\rm ^c}$ \\ 
\tableline {Mg} & 0.0005 & -0.0010 & 0 & 0.0260 & {0.0003 } & 
0.0271/0.0256 & 0.0361/0.0341 & - & - & - \\ 
{B} & -0.0004 & 0.0008 & 0 & 0.0027 & {-0.0007} & 0.0016/0.0028 & 
0.0024/0.0042 & 0.0175 & 0.006 & -0.0005%
\end{tabular}%
\end{table}
\begin{table}[tbp]
\caption[Tab. 2]{ Relaxation rate $1/T_{1}T$ in $10^{-3}$/(K sec). Both
unrenormalized and Stoner-enhanced values are included, as discussed in the
text. }
\begin{tabular}{c|cccccccccc}
& orbital & dipole & {Fermi-contact} & {core} & Total & Total (renormalized)
& Expt.\tablenote{Ref.\onlinecite{nmrosaka}; $^{\rm b}$ 
Ref.\onlinecite{nmrrussia};  ${\rm ^c}$ Ref.\onlinecite{nmrames}} & Expt.
$^{\rm b}$ & Expt. ${\rm ^c}$ &  \\ 
\tableline Mg & 0.02 & 0.01 & 1.0 & 0.0001 & 1.0 & 1.6 & - & - & - &  \\ 
B & 2.6 & 0.8 & 0.28 & {0.02} & 3.7 & 5.9 & 5.6 & 6.5 & 6.1 & 
\end{tabular}%
\end{table}
\begin{multicols}{2}

In order to understand the numerical results, we first calculate
analytically the shifts and the relaxation rate for a model Hamiltonian
which includes only B s and B p orbitals. We start with the contribution of
s electrons, i.e. the contact term. The contact shift may be written as $%
K\sim \mu _{B}^{2}(4/3)|\phi _{s}(0)|^{2}N_{ss}$, where $N_{ss}$ is the
s-projected DOS per atom per spin. We find $|\phi _{s}(0)|^{2}/(4\pi )\sim
2.16a_{0}^{-3}$, and $N_{ss}\sim 0.002$ states/eV per B atom. Therefore $%
K\sim 0.0026\%$ and $1/T_{1}T\sim 0.25\times 10^{-3}$ 1/(K sec). Both
numbers are very close to those obtained from the full calculations (Tables
1,2).

We now consider the contribution of B p electrons. The states at the Fermi
level are $\sim 70\%$ B p-like. $N_{p_{x},p_{x}}\sim 0.035$ states/(spin eV
atom), and $N_{p_{z},p_{z}}\sim 0.045$ states/(spin eV atom). Thus $%
N_{p_{x},p_{x}}\sim N_{p_{z},p_{z}}\sim N_{p}/3$, where $N_{p}$ is the total
p-projected DOS per spin per atom. Therefore we find the following
approximate expression of the orbital contribution to the relaxation rate

\[
\frac{1}{T_{1}T}\sim 4\pi k_{B}\hbar \gamma _{n}
^{2}\frac{4\mu _{B}^{2}}{3}|\langle r^{-3}\rangle
_{11}|^{2}Tr_{m}[{\bf l\cdot l}]\left( \frac{N_{p}}{3}\right) ^{2},\nonumber
\]
where $Tr_{m}[{\bf l\cdot l}]=(2l+1)l(l+1)=6$. We find $\langle
(a_{0}/r)^{3}\rangle \sim 1.14$ and $N_{p}/3\sim 0.038$ states/eV per B
atom, and therefore $1/T_{1}T\sim 3\times 10^{-3}$/(K sec). The orbital part
of the Knight shift is zero in this model because nondiagonal elements of
the DOS matrix vanish.

In the same way the dipole term can be written as

\[
\frac{1}{T_{1}T} \sim {8\pi k_{B}\hbar }\gamma _{n}
 ^{2}\mu _{B}^{2}|\langle r^{-3}\rangle _{11}|^{2} 
\sum_{\mu mm^{\prime }}\left( C_{1m,1m^{\prime }}^{2\mu }\right)
^{2}\left( \frac{N_{p}}{3}\right) ^{2}
\]
where $\sum_{\mu mm^{\prime }}(C_{1m,1m^{\prime }}^{2\mu })^{2}=6/5$. Thus
we find that the B p electron contribution to the dipole relaxation rate is $%
1/T_{1}T\sim 0.8\times 10^{-3}$ 1/(K sec). For the Knight shift we find $%
K_{z}^{d}\sim -2K_{xy}=2\mu _{B}^{2}\;C_{10,10}^{20}\langle r^{-3}\rangle
_{11}(N_{pp}/3)\sim 0.0015\%$. Again, all these numbers are rather close to
the all-electron results shown in the Tables. The ratio $%
(T_{1})_{dip}/(T_{1})_{orb}\sim (2/3)Tr[{\bf l\cdot l}]/\sum_{\mu mm^{\prime
}}\left( C_{1m,1m^{\prime }}^{2\mu }\right) ^{2}\sim 3.3$. The reason for
which the orbital term dominates over the dipole term is that all three $p$
are present at the Fermi level, as opposed, for instance, to the fullerenes%
\cite{Antropov}, where only one orbital is present at the Fermi level and
thus the orbital moment is quenched. 

In terms of the linear response theory, both the Knight shift and the
relaxation rate are defined by the electronic spin susceptibility\cite{moria}%
, $\chi ({\bf q},\omega {\bf )},$ specifically, $K\propto 
\mathop{\rm Re}%
\chi ({\bf 0},0{\bf ),}$ and $1/T_{1}\propto \lim_{\omega \rightarrow
0}\sum_{{\bf q}}%
\mathop{\rm Im}%
\chi ({\bf q},\omega )/\omega .$ Electron-hole excitations renormalized the
spin susceptibility, and in the simplest possible approximation one writes%
\[
\chi ({\bf q},\omega {\bf )\approx }\chi _{0}({\bf q},\omega {\bf )/[}%
1-I\chi _{0}({\bf q},\omega {\bf )],}
\]%
where $\chi _{0}$ is the bare (noninteracting) susceptibility, $I$ is the
so-called Stoner factor, characterizing intraatomic exchange, and the
calculations described above correspond to total neglect of the Stoner
renormalization. One can estimate $I$ from LSDA calculation with fixed total
spin moment by fitting the total energy to the Stoner expression, $%
E_{tot}(M)=M^{2}/4N-M^{2}I/4,$ where $M$ is the spin moment and $N$ is the
total DOS per spin. In this way, we found $IN\equiv I\chi ({\bf 0},0{\bf )}%
\approx 0.25.$ Thus we can estimate renormalized Knight shift as $K\approx
K_{0}/(1-IN)=1.33K_{0}.$ The renormalized values are also shown in the Tables.

Renormalization of $1/T_{1}$ is somewhat more difficult to take into
account. It is easy to show\cite{SA} that in the Stoner approximation%
\begin{equation}
\mathop{\rm Im}%
\chi ({\bf q},\omega {\bf )\approx }%
\mathop{\rm Im}%
\chi _{0}({\bf q},\omega {\bf )/[}1-I%
\mathop{\rm Re}%
\chi _{0}({\bf q},\omega {\bf )]}^{2}{\bf ,}
\label{imchi}\end{equation}
however, averaging this expression over ${\bf q}$'s requires knowledge of
the ${\bf q}$-dependence of $\chi _{0}.$ Generally speaking, renormalization
factor lies between 1$/(1-IN)$ and 1$/(1-IN)^{2}.$ Using the Lindhard
susceptibility ,and  a sphere for the Fermi surface, Shastry and Abrahams%
\cite{SA} found that in the 3D case 
\[
\left\langle \frac{%
\mathop{\rm Im}%
\chi _{0}({\bf q},\omega {\bf )}}{{\bf [}1-I%
\mathop{\rm Re}%
\chi _{0}({\bf q},\omega {\bf )]}^{2}}\right\rangle {\bf \approx }\frac{%
\left\langle 
\mathop{\rm Im}%
\chi _{0}({\bf q},\omega {\bf )}\right\rangle }{(1-IN)(1-2IN/3)},
\]%
which is a good approximation for $IN\lesssim 0.7.$ By integrating
numerically Eq.\ref{imchi} with the Lindhard function, we
found a better approximation, good for essentially all $IN,$ and preserving the
correct small $IN$ limit, namely  $\left\langle 
\mathop{\rm Im}%
\chi _{0}({\bf q},\omega {\bf )}\right\rangle /(1-IN)^{5/3}.$Thus we used
the factor 1.33$^{5/3}\approx $1.6 for $1/T_{1}.$ For the 2D free electron
gas, there is no {\bf q}-dependence in $\chi _{0}({\bf q},\omega {\bf )}$
for $q<2k_{F},$ and thus the renormalization factor is 1/$(1-IN)^{2}.$

The value of $1/T_{1}T=5.9\times 10^{-3}$ 1/(K sec), is in a good agreement
with the reported experimental number. This means that the DOS, calculated
within LDA, is a good approximation (maybe a slight underestimate) of the
bare DOS, and thus the values for the electron-phonon coupling constant $%
\lambda ,$ obtained from the specific heat measurements, are reliable.

To the best of our knowledge, there are at present no experimental data for
Mg. We predict that the magnetic shift is isotropic and that the principal
relaxation mechanism is the Fermi-contact interaction, despite of the fact
that $N_{ss}/\sum_{l>0}N_{ll}\sim 1/3$. The reason is that the quantities
that one has to compare are not the partial DOS $N_{ss}$ and $%
N_{ll}$ but rather the dimensionless couplings $(2\mu _{B}^{2}/3)|\phi
_{s}(0)|^{2}N_{ss}$ and $\mu _{B}^{2}\sum_{l>0}\langle r^{-3}\rangle
_{ll}N_{ll}$, and thus the relevant ratio is $R=(2/3)|\phi
_{s}(0)|^{2}N_{ss}/(\sum_{l>0}\langle r^{-3}\rangle _{ll}N_{ll})$. In the
case of Mg we find $|\phi _{s}(0)|^{2}/4\pi =4.54a_{0}^{-3}$, and $\langle
r^{-3}\rangle _{11}=4.8a_{0}^{-3}$. Hence we find $R\sim 5$. Instead, in the
case of B, $|\phi _{s}(0)|^{2}/4\pi =2.16a_{0}^{-3}$ and $\langle
r^{-3}\rangle _{11}=1.1a_{0}^{-3}$, and thus $R\sim 0.35$. The coupling with
non $s$ electron competes with or dominates over the coupling with $s$%
 electrons when $R\leq 1$

Finally, we would like to mention that the presented values for $1/T_{1}T$
include contributions from both quasi-2D $p_{\sigma }$ and 3D $p_{\pi }$
bands. If, as suggested\cite{multigap}, two different gaps open below $T_{c}$
in these bands, the temperature dependence of $1/T_{1}T$ $at$ low
temperature should be computed taking the different character of these bands
in the normal states. It is not obvious $apriori$ that the corresponding
weights will be just the densities of states. Calculations similar to those
described above, but band-decomposed are needed.

To summarize, we report first-principles calculations of the NMR relaxation
rates and the Knight shifts on both sites in MgB$_{2}.$ The results are in a
good agreement with the experiment, provided that the dipole and the orbital
hyperfine interactions are taken into account, as well as the Stoner
renormalization of susceptibility. NMR relaxation at $^{11}$B nucleus is
dominated by the orbital interaction, and that at the $^{25}$Mg nucleus by
the Fermi-contact one. The Knight shift is dominated by the Fermi contact
polarization both on B and on Mg. 
After these calculations were completed, we learned about similar
calculations  for the valence electrons relaxation rate on B 
from the Ames group\cite{Bel}, with the results consistent with
those reported here.

Useful discussions with E. Koch, O.K. Andersen, P. Carretta, V.P. Antropov,
and K.D. Belashchenko are gratefully acknowledged.

\end{multicols}

\end{document}